\documentstyle[epsfig,12pt]{article}
\def\today{\ifcase\month\or
  January\or February\or March\or April\or May\or June\or
  July\or August\or September\or October\or November\or December\fi
  \space\number\day, \number\year}

\def\deg{^\circ}

\def\ee{$e^+e^-$\ }
\def\ep{$e^-p$\ }

\def\cos{{\hbox{\rm cos}}}

\def\Leff{\Lambda_{\hbox{\it eff}}}
\def\frac#1#2{ {{#1} \over {#2} }}

\def\beq#1{\begin{equation}\label{#1}}
\def\beeq#1{\begin{eqnarray}\label{#1}}
\def\eeq{\end{equation}}
\def\eeeq{\end{eqnarray}}
\def\figcap{\section*{Figure Captions\markboth
        {FIGURECAPTIONS}{FIGURECAPTIONS}}\list
        {Figure \arabic{enumi}:\hfill}{\settowidth\labelwidth{Figure
999:}
        \leftmargin\labelwidth
        \advance\leftmargin\labelsep\usecounter{enumi}}}
 \relax

\def\CPC#1#2#3{Computer Phys. Comm. #1 (19#3) #2}
\def\PL#1#2#3{Phys.\ Lett.\ #1B (19#3) #2}
\def\NP#1#2#3{Nucl.\ Phys.\ #1B (19#3) #2}
\def\zp#1#2#3{Zeit.\ Phys.\ C#1 (19#3) #2}
\def\JL#1#2#3{JETP Lett.#1 (19#3) #2}
\def\JP#1#2#3{J.\ Phys.\ G#1 (19#3) #2}
\def\NIM#1#2#3{Nucl.\ Instr.\ Meth. A#1 (19#3) #2}
\def\PR#1#2#3{Phys.\ Rep.\ #1 (19#3) #2}
\def\RMP#1#2#3{Rev.\ Mod.\ Phys. #1 (19#3) #2}
\begin{document}
\begin{titlepage}
\begin{flushleft}
%
%
{\tt DESY 95-072, hep-ex/9505003    \hfill    ISSN 0418-9833} \\
{\tt April 1995}                  \\
\end{flushleft}
\vspace*{4.cm}
\begin{center}
\begin{Large}
{\bf  A Study of the Fragmentation of Quarks} \\
{\bf{in} \bf\boldmath{$e^-p$} {\bf Collisions at HERA}} \\
\vspace*{2.cm}

H1 Collaboration \\
\end{Large}
\vspace*{4.cm}
{\bf Abstract:}
\begin{quotation}
         Deep inelastic scattering (DIS) events, selected from 1993 data taken
by the H1 experiment at HERA, are studied in the Breit frame of reference.
The fragmentation function of the quark is compared with those of \ee data.
It is shown
that certain aspects of the quarks emerging from within the proton in \ep
interactions are essentially the same as those of quarks pair-created from the
vacuum in \ee annihilation. The measured area, peak position and width of the
fragmentation function show that the kinematic evolution variable, equivalent
to the \ee squared centre of mass energy, is in the Breit frame the invariant
square of the four-momentum transfer. We comment on
the extent to which we have evidence for coherence effects in parton showers.
\end{quotation}
\vfill
%
%
\end{center}
\end{titlepage}
\noindent
{
\noindent
S.~Aid$^{13}$,                   
 V.~Andreev$^{24}$,               
 B.~Andrieu$^{28}$,               
 R.-D.~Appuhn$^{11}$,             
 M.~Arpagaus$^{36}$,              
 A.~Babaev$^{26}$,                
 J.~Baehr$^{35}$,                 
 J.~B\'an$^{17}$,                 
 Y.~Ban$^{27}$,                   
 P.~Baranov$^{24}$
% ...

\noindent
% ...
$ ^{11}$ DESY, Hamburg, Germany$^a$ \\
% ...
$ ^{13}$ II. Institut f\"ur Experimentalphysik, Universit\"at Hamburg,
                                     Hamburg, Germany$^ a$  \\
% ...
$ ^{17}$ Institute of Experimental Physics, Slovak Academy of
                Sciences, Ko\v{s}ice, Slovak Republic$^ f$\\
% ...
$ ^{24}$ Lebedev Physical Institute, Moscow, Russia$^ f$ \\
% ...
$ ^{26}$ Max-Planck-Institut f\"ur Physik,
                                            M\"unchen, Germany$^ a$\\
 $ ^{27}$ LAL, Universit\'{e} de Paris-Sud, IN2P3-CNRS,
                            Orsay, France\\
 $ ^{28}$ LPNHE, Ecole Polytechnique, IN2P3-CNRS,
                             Palaiseau, France \\
% ...
$ ^{35}$ DESY, Institut f\"ur Hochenergiephysik,
                              Zeuthen, Germany$^ a$\\
 $ ^{36}$ Institut f\"ur Teilchenphysik,
          ETH, Z\"urich, Switzerland$^ i$\\
% ...
$ ^a$ Supported by the Bundesministerium f\"ur
                                  Forschung und Technologie, FRG
 under contract numbers 6AC17P, 6AC47P, 6DO57I, 6HH17P, 6HH27I, 6HD17I,
 6HD27I, 6KI17P, 6MP17I, and 6WT87P \\
% ...
$ ^e$ Supported in part by USDOE grant DE F603 91ER40674\\
 $ ^f$ Supported by the Deutsche Forschungsgemeinschaft\\
% ...
$ ^i$ Supported by the Swiss National Science Foundation\\

}
\vskip 2mm
\noindent  {\bf 1. Introduction} \vskip .1 true cm

\noindent This paper analyses the spectra of hadrons produced in deep
inelastic scattering (DIS) \ep collisions as a function of both
the four-momentum squared ($q^2=-Q^2$) transferred from the electron and the
Bjorken scaling variable, $x$. In terms of the proton four
momentum $P$, $x$ is defined as $Q^2/2P\cdot q$, but it may be na\"\i vely
thought of as the momentum fraction of the struck quark in the proton.
We also select on the basis of the dimensionless inelasticity variable
$y= Q^2/xs$ where $\sqrt{s}$ is the overall \ep centre of mass
energy.
 The H1 collaboration has already published charged particle
spectra, analysed in the hadronic centre of mass (CMS) frame, in intervals of
the total hadronic mass, $W$ (where $W^2\approx Q^2(1/x-1)$), using
22.5~nb$^{-1}$ of data obtained in the first year of HERA operation
[\ref{H1spec}]. It is the aim of the present analysis to
use the increased statistics of 1993 data to cast such spectra
into the form of fragmentation functions which are reliably related to the
hadronisation of quarks. Aspects of these distributions can then be used in a
comparison with corresponding data from \ee experiments [\ref{OPAL}]
to determine the relevant evolution variable for such distributions.

  Since the initial state of \ee annihilation interactions is a
neutral off mass-shell photon or Z$^0$,  all events with only hadrons in the
final state are the result of a pure quark-antiquark pair creation process.
Hard interactions in $pp$ or $p\bar p$ collisions are much more complicated.
Apart from the effect of the proton-remnant hadronisation, initial state QCD
radiation predominantly involving gluons is important. In addition gluon-gluon
interactions dominate the hard scatter. Quark
identification can only be performed statistically and in very limited
kinematic areas. In \ep interactions at HERA there are fewer problems caused by
the proton remnant and, to first order, the virtual photon couples only to
charged quarks. Electron kinematics alone can give the final state quark
kinematics thus enabling a study of its hadronic fragments for direct
comparison with quark fragmentation in \ee interactions.

        The momentum distribution of the hadrons from quark fragmentation in
\ee interactions roughly scales with $\sqrt{s_{ee}}=E^*$,
the overall centre of mass system (CMS) energy. The ratio of the momentum
of a given charged hadron to the maximum energy that it could have,
$x_p=2p^{\pm}_{hadron}/E^*$, is thus a natural variable in terms of which \ee
experiments [\ref{OPAL},\ref{Tasso}] have described hadronic spectra.
It has the advantage of being independent of the difficulties involved in any
jet or thrust axis determination. 
The variable is manifestly not Lorentz invariant and it will therefore become
necessary to identify an equivalent frame for \ep interactions in which
comparisons can properly be made.

        The distribution
$D^{\pm}(x_p) = (1/N_{evts})\times dn^{\pm}_{tracks}/dx_p$, being
a charged track density normalised by the number of events, is termed the
fragmentation function. It characterises the complete process
which includes parton shower development as well as non-perturbative
hadronisation. In principle the fragmentation function is unique only to a
given quark and hadron species, but tagging techniques usually dictate that
the data is effectively an integrated average.
 It is a ``soft'' function rising rapidly as $x_p$ decreases
but then turning over near $x_p=0$. A related distribution is that of the
arbitrarily normalised logarithm of track energies where the turn-over
becomes the ``hump-backed'' plateau [\ref{Hump}].

The variable $\xi=\ln (1/x_p)$ expands the turn-over region, and the
expectation
of the Modified Leading Logarithmic Approximation (MLLA) to perturbative QCD
predicts that for partons $\xi$ is distributed in a roughly Gaussian
manner. Assuming Local Parton-Hadron Duality (LPHD) the same
behaviour is expected for any type of hadron with the mean $\xi$
increasing as  $\ln (E^*)$ and the area under the Gaussian (multiplicity)
increasing only slightly faster [\ref{Dok}].
More detailed expectations of the fragmentation stem from the influence of
quantum-mechanical coherence between gluons in the time-like parton shower
[\ref{Coh}].
Effectively this reduces the available phase space for soft gluon emission to
an angular ordered region. This coherence should then reduce the
gradient of the energy dependence of the peak position.

        In \ep interactions there is no exact equivalent of the \ee
centre of mass system. The hadronic ($\gamma p$) CMS is one candidate. In the
na\"\i ve quark parton model (QPM) the final
state then contains a quark which is back to back with the spectator system.
This collaboration has already published [\ref{H1spec}]
$x_F= p_z/p_{max}$ distributions in this frame for the hemisphere opposite to
the direction of travel of the incoming proton and outgoing remnant.

      A further boost along the $z$-(virtual photon)axis gives the Breit
frame defined in this analysis such that the positive $z$-direction is that
of the incoming quark and the negative direction that of the incoming photon.
This frame is defined such that the virtual photon is entirely space-like,
having a momentum of $p_z=-Q$. Within the QPM a quark has momentum $p_z=+Q/2$
before and $p_z=-Q/2$ after scattering and therefore, in this frame,
$x_p=2p^{\pm}_{hadron}/Q$ is the equivalent of the
\ee definition of a fragmentation variable\footnote{Note that since $Q^2$ is
defined as a positive quantity one can refer to a value of `$Q$' for the
event. References [\ref{BF}] discuss other possible variables in the Breit
frame incorporating, for example, $p_{\parallel}$ which gives directional
information. Since the main aim of this paper is
to compare with \ee experiments, we have chosen to remain with a compatible
definition.}.  QCD radiation alters these values and implies that the collision
is no longer collinear. In particular initial-state radiation
of gluons has no equivalent in an \ee event.
Note that whereas a given \ee experiment has one fixed
value of $E^*$, an \ep experiment results in a wide range of $Q$ enabling
evolution of the fragmentation function to be tested in a
single experiment. Conversely the fragmentation function should now be
considered to be the result of a convolution with the structure functions of
the individual quarks. This analysis will ignore this effect but will look
for variations of fragmentation as a function of the event $x$ as well
as $Q^2$.



        The Breit frame is defined solely by the kinematics of the virtual
current and is concerned more with the quark-current rather than the
proton-current interaction. The spectator system has reduced significance.
Thus, despite some difficulties [\ref{GTetal}],
the negative $z$ hemisphere of this frame, rather than the $\gamma p$ CMS,
has been
suggested [\ref{Dok}] as a better approximation to one half of the \ee CMS.
Monte Carlo comparisons have been made [\ref{Char},\ref{Ingel}] to indicate the
possibility of making the kind of measurements presented here. These analyses
merely {\it define} a given hemisphere to contain current or
target-associated tracks but in this analysis we shall also make some attempt
to assess the confidence with which such associations may be made.
Quantum-mechanically there is no absolute meaning for such associations,
but as with the existence of jets, there are useful approximations
which can be made. Since the $\gamma p$ CMS and the Breit frames are linked
by a boost, they cannot in general both have the same negative $z$
hemisphere content so the question of relative loss and contamination is
worthy of empirical examination.

\noindent  {\bf 2. The H1 Experiment} \vskip .1 true cm

\noindent  This analysis uses data from 1993  when the HERA
\ep collider was run with a proton beam of 820~GeV and an electron beam of
26.7~GeV. A detailed description of the H1-detector has been given elsewhere
[\ref{H1det}]. This paper will thus give only a short description of
components vital for this analysis.

        Momentum measurements of charged tracks are provided, in the central
region, by two cylindrical and co-axial drift chambers [\ref{H1CDT}] for
($r,\phi$) measurement supplemented by two $z$-chambers; in the forward
(proton) direction they are provided by three Radial and three Planar drift
chamber modules [\ref{FTD}]. These track detectors are inside a uniform
1.15~T magnetic field. Track segments from all devices are combined to give
efficient
detection and momentum measurement with $\delta p/p^2$ better than 1\% /GeV
in the angular range used in this analysis, $10\deg <\theta<160\deg $.
A full Monte Carlo simulation 
based on GEANT [\ref{GEANT}] gives an acceptable description of the effect of
dead areas and reconstruction efficiency.

        In the polar angle range $4\deg <\theta<153\deg $ the trackers
are surrounded by a fine-grained liquid argon (LAr) sampling calorimeter
[\ref{H1CAL}] with lead and steel absorbers in the electromagnetic and
hadronic sections, respectively. The calorimeter cells measure hadronic energy
flow and the energy of the scattered electron for high $Q^2$ events. The LAr
calorimeter is complemented by a backward electromagnetic, lead-scintillator,
calorimeter (BEMC) covering the angular range $151\deg <\theta<176\deg $.
The data are divided into
two separate samples for low and high $Q^2$ where the scattered electron
energy is well contained in either but not both of the two calorimeters. There
are possible systematic energy scale errors taken to
be $\pm$1.7\% for the BEMC and $\pm$3\% for the LAr calorimeter.

\newpage

\noindent  {\bf 3. Data Selection and Acceptance Corrections} \vskip .1 true cm

\noindent  The source data for this analysis come
from the late period of 1993 when magnetic field was reestablished in the H1
detector. Selections to further limit data only to periods when
there were extremely stable tracking conditions and when there was no
coincident problem with electronic `noise' in the H1 calorimetry give a total
integrated luminosity for this subset of $\approx 150$~nb$^{-1}$.

        A neutral current DIS event selection is made by demanding a
well-reconstructed scattered electron of energy greater than 14~GeV. The
selection $12<Q^2<80$ GeV$^2$ defines the BEMC sample,
whereas the requirement for the scattered electron to have a polar angle in the
range $10\deg<\theta<150\deg$ (resulting in the constraint $Q^2>100$ GeV$^2$)
defines the LAr sample. These data sets are referred to as the low and high
$Q^2$ samples, respectively. In terms of the usual DIS variables further cuts
of $W^2>3000$~GeV$^2$ and, in the case of high $Q^2$, $y<0.6$ are applied to
maintain high acceptance, and to avoid radiative corrections and contamination
from photoproduction events. The final sample is better than 99.7\% pure DIS.
Additional cuts [\ref{H1spec}] are made to exclude possible diffractive events
(about 6\% of the sample) where the virtual photon may be said to interact
with a Pomeron-like object [\ref{diff}]. There is no equivalent in
the \ee interactions with which we intend to compare and such events are not
adequately described by our DIS Monte Carlo generators with which we perform
acceptance corrections. In practice this removal is done by demanding that
there is at least 0.5~GeV deposit of observable energy in the polar region
$4.4\deg<\theta<15\deg$. The resultant event numbers are given as
the first column in Table \ref{EvData}.

\begin{table}
\begin{center}
\begin{tabular}{|l|l|l|l|l|}
\hline
Data                   & Non diffractive & Good central & Breit frame    \\
Sample                 & events          & calorimetry  & energy flow  \\
\hline\hline
$12<Q^2<80$ GeV$^2$    &\hfill 5927     &\hfill  3141 &\hfill  1945   \\
\hline\hline
$Q^2>100$ GeV$^2$      &\hfill  423     &\hfill   373 &\hfill   235   \\
\hline
\end{tabular} \end{center}
\caption{ \sl Size of data samples for this analysis. See text for definition
of
selections.}
\label{EvData}
\end{table}

        Within these events, charged tracks are selected with
$\delta p/p<1$ and with transverse momentum $p_t>150$~MeV/$c$ in the central
tracker or momentum $p>500$~MeV/$c$ in the forward tracker. The tracks are
also required to satisfy basic quality criteria on total number of
digitisations, quality of vertex fits etc. Together these
selections give a smooth variation in acceptance over the whole angular range.
Cuts are made to remove tracks not originating from the
primary interaction vertex but these fail to exclude roughly half
of the $\approx 4$\% of tracks arising from $K^0_s$ decays.

        Corrections for the effect of all of these cuts, and for the
contribution of remaining $K^0$ and $\Lambda$ decay products, are
made by reconstructing Monte Carlo events with the same code and
selections and then comparing these with results for primary charged
particles at the generated event level.  Commonly, LEPTO 6.1 [\ref{H1spec}]
is used with
MRS~H structure functions and with a Colour Dipole model [\ref{CDM}] for
gluon radiation, but a version with matrix elements matched to parton showers
in the MLLA approximation [\ref{Lepto}] has also been used and we note no
significant change in any of the results. Since corrections are made
bin-by-bin, in narrow kinematic areas, and with {\it ratios} of generated to
reconstructed track numbers the analysis is rather insensitive to Monte Carlo
physics assumptions.
%
%

\noindent  {\bf 4. Breit Frame Event Selection} \vskip .1 true cm

\noindent        It is the aim of this analysis to select well-measured
events requiring small acceptance corrections for analysis of tracks in
the Breit frame of reference.  This may be done using calorimetery
information alone in order to avoid track reconstruction bias. First
four-momentum vectors which correspond to the direction and energy of
calorimeter clusters (assuming pion masses) are boosted
to the Breit frame using the ($x,Q^2$) values calculated solely from
the electron kinematics.
The $z<0$ hemisphere, commonly referred to as the current fragmentation region,
is then examined in more detail. To ensure that events have good and uniform
track acceptance a selection is made such that events are taken for further
analysis if 95\% or more of the observed energy in this hemisphere originates
from the laboratory polar region $10\deg<\theta<150\deg$. This region is
completely covered by the (LAr) calorimetry and thus has consistent
calibration. The numbers of such events surviving this cut are also given in
Table \ref{EvData} in the column headed `good central calorimetry'.

\begin{figure}
 \begin{center}
  \mbox{\epsfig{file=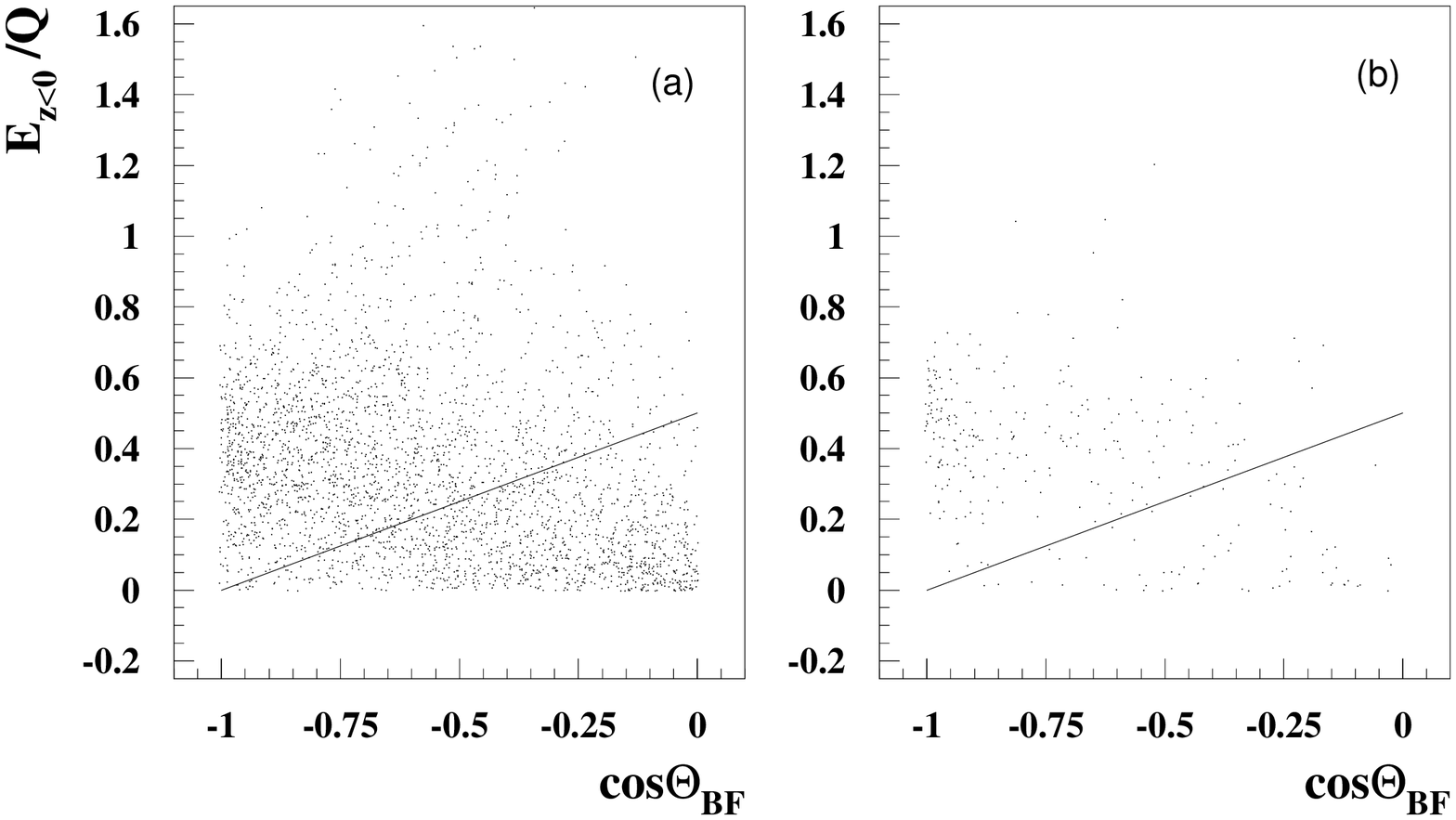,height=9.0cm}}
 \end{center}
 \caption{\em The total energy of the summed calorimeter cluster four momentum
vectors in the $z<0$ hemisphere of the
Breit frame is plotted as a fraction of the event $Q$ against the cosine of
the polar angle of the
resultant vector, for (a) the low $Q^2$ and (b) the high $Q^2$ sample.
The line indicates the cut for the secondary analysis described in the text.
}
\label{ECSTH}
 \begin{center}
  \mbox{\epsfig{file=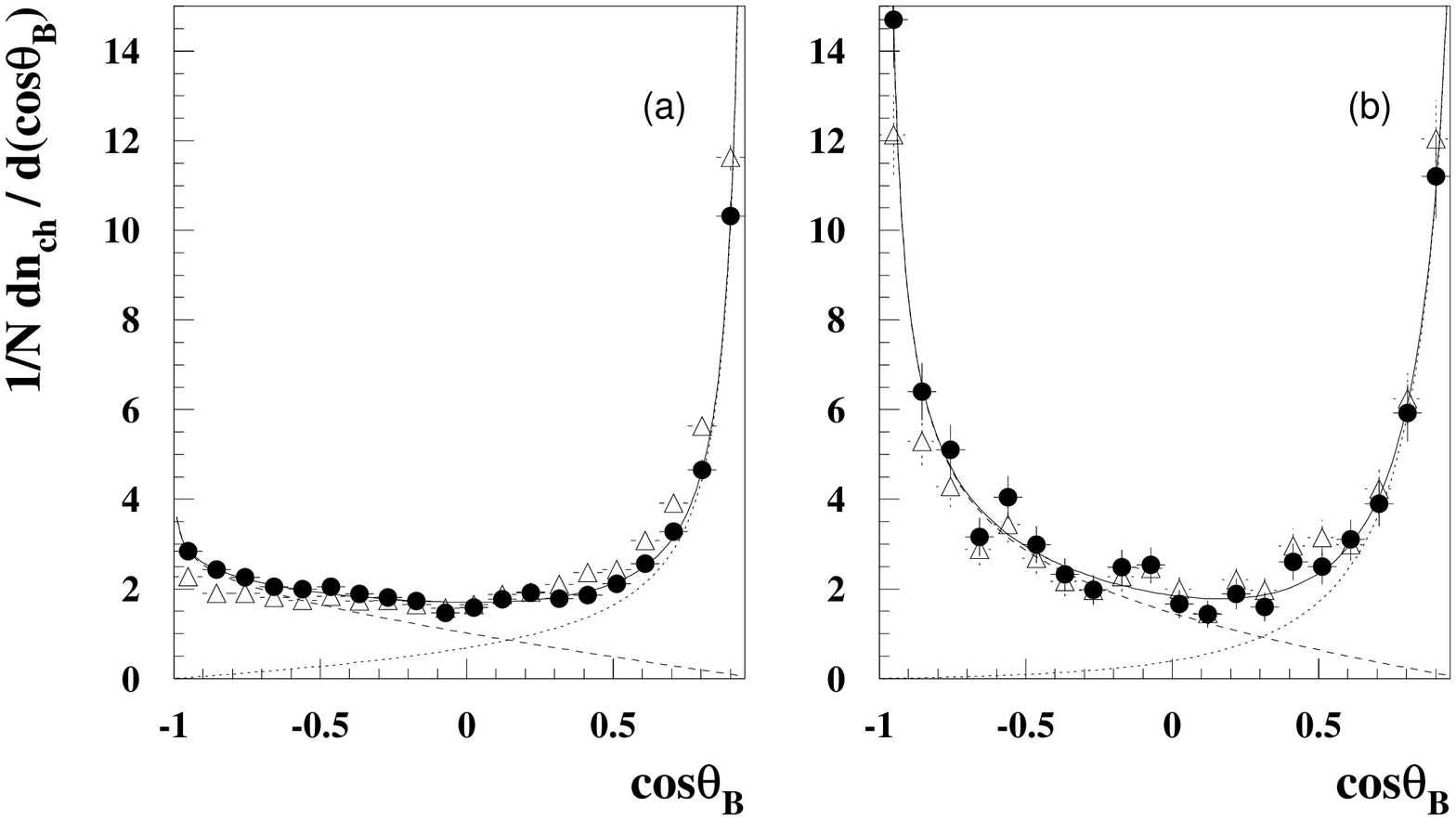,height=9.0cm}}
 \end{center}
 \caption{\em Distribution of the cosine of the Breit frame polar angle for
tracks of (a) the low $Q^2$ and (b) the high $Q^2$ sample, with statistical
errors only.
The open triangles show the data before the Breit frame energy flow selection.
The solid line corresponds to the empirical fit described in the text,
where the dashed line is the nominal quark contribution and the dotted line
that of the target.
}
\label{Cost}
\end{figure}

          For each event, the four-momenta of all energy clusters in the
$z<0$ hemisphere are added and the resultant total energy $E_{z<0}$, is
plotted as
a fraction of the event $Q$ against the resultant cosine of the Breit frame
polar angle, $\cos \Theta_{BF}$, in Fig.~\ref{ECSTH}. According to the QPM, a
quark would be expected to have energy $Q/2$ and $\cos \Theta_{BF} = -1$. At
high $Q^2$ there is a tendency for the events to cluster in the area of this
point. However, the spread of events is much greater at low $Q^2$ and there
are a large number of events with small energy in the `current' hemisphere
and/or with a net cluster emerging at a large angle with respect to the
beam axis. If, for a given event, the net energy flow is only just
within the current hemisphere there could be some risk of bias in measuring the
share of that energy. Simulated events samples behave in exactly the same way
as the data and studies have shown that this spread is mainly because of
reconstruction effects. Simple rejection of such events is a little
problematic however as, at the generated level, some of these events show
slightly different properties because of the loss of tracks to the target
hemisphere through significant QCD radiation some of which would also have its
equivalent in \ee interactions. We thus choose to test the possibilities of
bias by performing  the analysis both on all events and also by rejecting
events below the line joining $E_{z<0}=0$, $\cos\Theta_{BF}=-1$ to
$E_{z<0}=Q/2$, $\cos\Theta_{BF}=0$ indicated in
Fig.~\ref{ECSTH} and referred to as the Breit frame energy flow selection.
The event numbers surviving this topological selection are given as the final
column in Table \ref{EvData}.

        Having selected events using calorimetry, {\it track} properties
in the Breit frame are now investigated, beginning with the distribution of
$\cos\theta_B $, the cosine of the polar angle for charged particles, as shown
in Fig.~{\ref{Cost}}(a,b).

        These data have bin-by-bin corrections for losses due to acceptance
and inefficiencies calculated using simulated and reconstructed
Monte Carlo events. For $-1<\cos\theta_B < 0.99$, avoiding effects in this
frame due to the beam pipe, the average correction factor is 1.35 for the low
$Q^2$ sample and 1.42 for the high $Q^2$ sample. The correction factors are
small and vary smoothly for $\cos\theta < 0.0$. There is an obvious peak in the
positive-$z$ (beam) direction and also in the negative-$z$ (quark)
direction at high $Q^2$ but which is much less clear at low momentum
transfer.

  In strict quantum-mechanical interpretation, no track can be said to
`belong' to either the jet or target remnant, wherever it lies in the
Breit frame. It is clear from Fig.~\ref{Cost} however that the concept of
current and target hemispheres is a good approximation as $Q^2$ increases
but almost a question of mere definition at our lowest $Q^2$ values. We adopt
two different procedures to estimate the possible loss or contamination
involved in using a selection of one hemisphere of the Breit frame to represent
the quark behaviour.
In Fig.~\ref{Cost} there are empirical fits to
$Ae^{-a(1+x)^\alpha}(1-x) + Be^{-b(1-x)^\beta}(1+x)$, where $x=\cos\theta_B$.
This and fits to other possible functional forms suggest that, in the Breit
frame, the selection $\cos\theta_B <0.0$ roughly equalises loss and
contamination between the current and target fragmentation and that they are
at the $<5$\% level at high $Q^2$ but at the $\approx 20-25$\%
level at low $Q^2$.
Our method of correction by Monte Carlo simulation accounts for migration
between hemispheres caused by resolution effects. A separate study showed that
these are dominated by measurement errors of the reconstructed electron giving
an incorrect boost rather than track measurement errors. The magnitude of
migration losses and contaminations are commensurate with the various
empirical fits suggesting that these latter are indeed dominated by
poor reconstruction.
It is the approach of this analysis to adopt $\cos\theta_B <0.0$ as our
selection for tracks.  If it is further required to associate these
particle multiplicities with a quark  we feel that an additional
20(5)\% systematic error on the particle multiplicity of low(high) $Q^2$ data
should be imposed to describe our remaining ignorance of current/target
separation.

\noindent  {\bf 5. Fragmentation Functions} \vskip .1 true cm

\begin{figure}
 \begin{center}
 \mbox{\epsfig{file=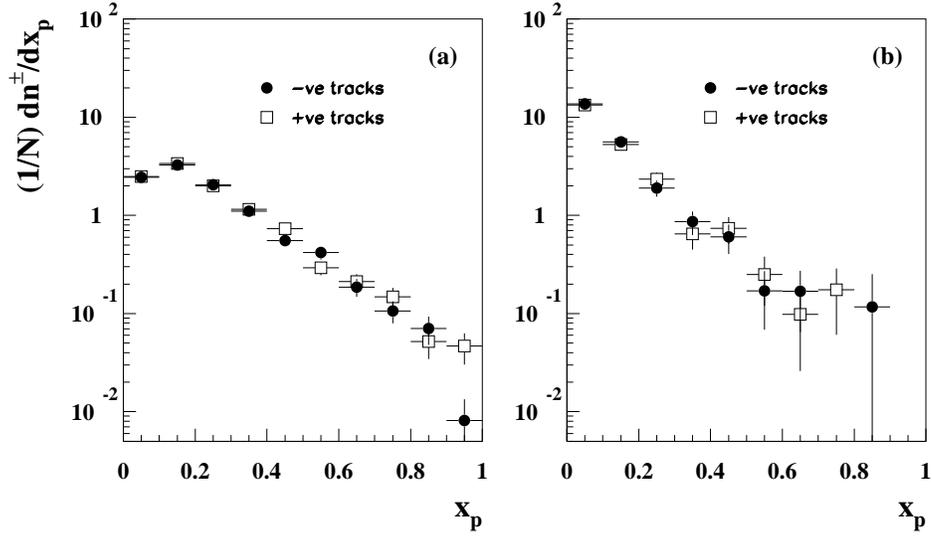,height=9.0cm}}
 \end{center}
 \caption{\em The fragmentation functions, $D^{\pm}(x_p)$, for the current
hemisphere of the Breit frame shown separately for positive and negative
tracks, for (a) the low $Q^2$ and (b) the high $Q^2$ sample with statistical
errors only.
}
\label{Dx}
\end{figure}

\begin{figure}
 \begin{center}
 \mbox{\epsfig{file=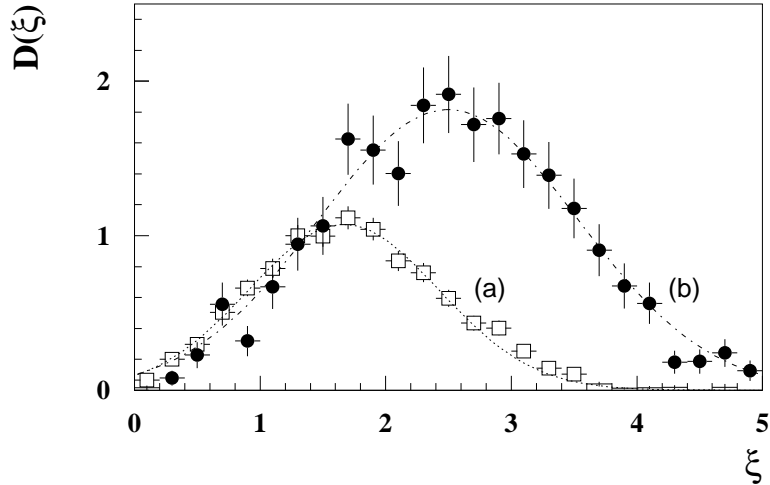,height=9.0cm}}
 \end{center}
 \caption{\em The fragmentation functions for the current hemisphere of the
Breit frame, $D(\xi)$, for
(a) the low $Q^2$ and (b) the high $Q^2$ sample, with statistical errors only
and with simple Gaussian fits superimposed.
}
\label{Dxi}
\end{figure}

\noindent  The charged particle fragmentation functions, as defined in the
introduction, are displayed in Fig.~\ref{Dx}(a,b) as a function of
$x_p=2p^{\pm}_{hadron}/Q$ for the current hemisphere of the Breit frame and
for the two $Q^2$ intervals.  These figures utilise the selections of the
previous section including the Breit frame energy flow selection.
The individual fragmentation functions, for positive and negative hadrons,
are indistinguishable, at least with the present statistics in the current
kinematic areas of investigation, and are treated inclusively for the rest of
this paper. Changing the fragmentation variable to be $\xi=\ln (1/x_p)$ and
defining the fragmentation function to be
$D(\xi) = (1/N_{evts})\times dn_{tracks}/d\xi$ results in the Gaussian shape
of Fig.~\ref{Dxi}(a,b) expected from the discussion in section~1 and seen also
in \ee data. The area increases and the peak  moves to higher values of $\xi$
at higher $Q^2$.
These distributions are also corrected
bin-by-bin for acceptance and efficiency, although again the dependence is
smooth and small\footnote{
The effect of remaining QED radiative corrections to the Born term has been
investigated using the Monte Carlo program Django 2.1 [\ref{Djn}]. This
analysis shows no effective variation over the peak area but an overall
normalisation correction of $+4.3\pm 0.3$\% has been included within our
corrections, although it is small compared with other sources of
error.} especially in the case of the variable $\xi$.
Simulation studies show that the resolution in $\xi$ ($\delta\xi\approx 0.12$),
again dominated by the uncertainties in the boost to the Breit frame is
smaller than the bin-width of 0.2 in $\xi$ used throughout in this analysis.

\begin{figure}
 \begin{center}
 \mbox{\epsfig{file=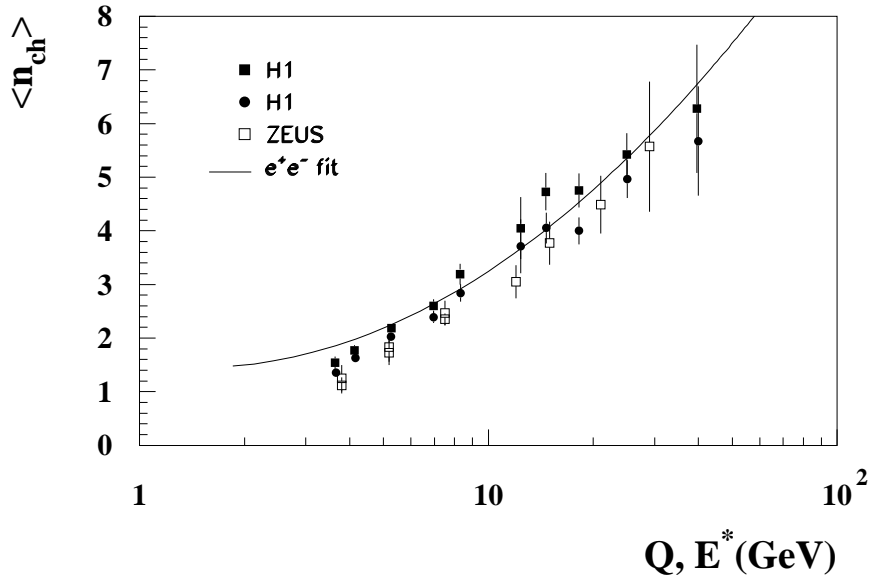,height=10.0cm}}
 \end{center}
 \caption{\em Average charged particle multiplicity in the current region of
the Breit frame for data before (solid circles) and after (solid squares) the
energy flow selection as a
function of $Q$ compared with a parameterisation
(line) of one half of average corrected track multiplicities in \ee
events as a function of $E^*$. The results of a ZEUS analysis are shown as open
squares, with points at the same $Q$ being at different $x$.
}
\label{Mult}
\end{figure}

The low and high $Q^2$ data of Fig.~\ref{Dxi} have each been further subdivided
to give a total of ten intervals of $Q$ in order to study the evolution of
the area, peak and width of the fragmentation function in more detail.
The variation of the integrated area (equal to the mean charged particle
multiplicity, $<n_{ch}>$) is shown in Fig.~\ref{Mult} and the results are
reproduced in Table \ref{TMult}.
The systematic errors described at the end of section~4 dominate but are added
in quadrature with statistical errors. The figure and table show the results
obtained
both with and without the Breit frame energy flow selections described in
section~4 since there appears to be a significant difference, especially at
low $Q^2$. Since these data are corrected for known acceptance effects we
ascribe this difference to QCD radiation depositing particles in the target
hemisphere. It is seen similarly at the generated level of Monte Carlo. The
difference is indicative of difficulties of detail inherent in a comparison
with
\ee data. A fit [\ref{eemult}] to many \ee
data points\footnote{Note that the published results refer to \ee
{\it events}. Average track multiplicities have been reduced by a factor two
to correspond to results for each {\it quark} and a further 8.1\% (the
average of available data) reduction has been made to account for
$K^0$ and $\Lambda$ decay tracks.}
is also shown for comparison as a function of $E^*$. At low $Q$
our results, and those of the ZEUS experiment [\ref{Zeus}], comparable to the
H1 results with no selection, show a significantly lower average charged
multiplicity but compatible with differences that might be expected from
differing QCD radiation effects.

\begin{table*}
\begin{center}
\begin{tabular}{|c||c|c|c||c|c|c|} \hline
\multicolumn{1}{|c||}{\bf {Q$^2$ Interval}} &
\multicolumn{3}{|c||}{\bf {Energy Flow Selection}} &
\multicolumn{3}{|c|}{\bf {Total Current Hemisphere}} \\ \hline \hline
\multicolumn{1}{|c||}{ (GeV$^2$)}
&\multicolumn{1}{|c|}{Multiplicity}
&\multicolumn{1}{|c|}{Peak}
&\multicolumn{1}{|c||}{Width}
&\multicolumn{1}{|c|}{Multiplicity}
&\multicolumn{1}{|c|}{Peak}
&\multicolumn{1}{|c|}{Width} \\ \hline \hline
{12 $\to$ 15} &
{1.54$\pm$0.11}&{1.41$\pm$0.13}&{0.83$\pm$0.14}&
{1.35$\pm$0.08}&{1.45$\pm$0.08}&{0.72$\pm$0.08} \\ \hline \hline
{15 $\to$ 20} &
{1.77$\pm$0.10}&{1.47$\pm$0.10}&{0.72$\pm$0.07}&
{1.63$\pm$0.08}&{1.49$\pm$0.08}&{0.82$\pm$0.09} \\ \hline \hline
{20 $\to$ 40} &
{2.18$\pm$0.08}&{1.63$\pm$0.09}&{0.71$\pm$0.04}&
{2.02$\pm$0.06}&{1.66$\pm$0.06}&{0.74$\pm$0.04} \\ \hline \hline
{40 $\to$ 60} &
{2.60$\pm$0.13}&{1.91$\pm$0.11}&{0.83$\pm$0.10}&
{2.39$\pm$0.11}&{1.94$\pm$0.09}&{0.91$\pm$0.11} \\ \hline \hline
{60 $\to$ 80} &
{3.20$\pm$0.19}&{2.00$\pm$0.12}&{0.83$\pm$0.12}&
{2.84$\pm$0.16}&{2.01$\pm$0.10}&{0.87$\pm$0.13} \\ \hline \hline
{100 $\to$ 175} &
{4.05$\pm$0.58}&{2.23$\pm$0.39}&{1.00$\pm$0.71}&
{3.71$\pm$0.50}&{2.27$\pm$0.17}&{0.75$\pm$0.27} \\ \hline \hline
{175 $\to$ 250} &
{4.73$\pm$0.35}&{2.12$\pm$0.34}&{1.28$\pm$0.50}&
{4.05$\pm$0.29}&{2.18$\pm$0.30}&{1.26$\pm$0.48} \\ \hline \hline
{250 $\to$ 450} &
{4.75$\pm$0.32}&{2.72$\pm$0.29}&{1.24$\pm$0.45}&
{4.00$\pm$0.25}&{2.68$\pm$0.23}&{1.09$\pm$0.32} \\ \hline \hline
{450 $\to$ 1000} &
{5.43$\pm$0.39}&{2.70$\pm$0.13}&{0.80$\pm$0.15}&
{4.97$\pm$0.36}&{2.71$\pm$0.16}&{0.79$\pm$0.17} \\ \hline \hline
{1000 $\to$ 8000} &
{6.27$\pm$1.20}&{2.62$\pm$0.28}&{0.88$\pm$0.39}&
{5.67$\pm$1.02}&{2.63$\pm$0.26}&{0.85$\pm$0.35} \\ \hline \hline
\end{tabular}
\end{center}
\caption{ \sl Average charged particle multiplicity, peak and width
of the fragmentation function for the $Q^2$ intervals
given using either the energy flow selection or the total current
hemisphere of the Breit frame.}
\label{TMult}
\end{table*}

At a detailed level, agreement between  particle multiplicities
can also not be expected to be precise because of the comparative `flavour
democracy' expected in \ee annihilation. Jets initiated by
$b$-quarks, constituting some
22\% of \ee events, have roughly 13\% higher average charged multiplicity at
LEP energies [\ref{Delphi}]. Such flavour effects are thus at the 3\% level
which is much less significant than our present errors and we make no attempt
to correct for them.
There is no doubt from this figure that $Q$ is a suitable scaling
quantity equivalent to $E^*$ in \ee interactions.
Low energy, fixed target data [\ref{EMC1}] indicated agreement between
average charged particle multiplicities in the current hemisphere of the
hadronic CMS frame and those of \ee with $E^*$ at the same $W$ but later
experiments [\ref{H1spec},\ref{EMC665}] show significant disagreements at
low Feynman-$x$. These analyses do not, in any case, attempt
to demonstrate the relatively clear current-target separation
seen, at least at high $Q^2$, in this analysis.

The peak and width ($\sigma$) values of the fragmentation function may be
found from Gaussian fits to the central area ($\pm$ one unit of $\xi$
either side of the statistical mean) so as simultaneously to test MLLA
predictions and to minimise dependence on Monte Carlo corrections which
mainly affect the tails of the distributions.
The simulated event studies referred to at the end of section~4 show that
the resolution of a particle's polar angle in the Breit frame is much poorer
than its momentum and indicate that at $\cos\theta_B\approx 0$,
$\delta(\cos\theta_B)\approx\pm$ 0.40(0.25) at low(high)
$Q^2$. This quantity does not enter into $\xi$ but affects peak and
width measurements through uncertainty in the selection of the
negative Breit frame hemisphere. As an estimator of the systematic error that
might be introduced because of this selection we have repeated the above
analyses using the cuts at the $\pm$ one sigma level of these resolutions and
use extreme differences of results as an estimate of systematic errors.
For the peak values, this estimate gives a result which is, in general, of
the order of the statistical fitting errors. There is an additional source of
systematic error of $\approx\pm$6\% in $\xi_{peak}$ which arises from
uncertainty in the boost due to the calorimeter energy scale errors given
at the end of section~2. These errors dominate all other
sources of systematic error coming, for example, from using different Monte
Carlo generators for corrections.
Width measurements are completely dominated by statistical errors.

\begin{figure}
 \begin{center}
  \mbox{\epsfig{file=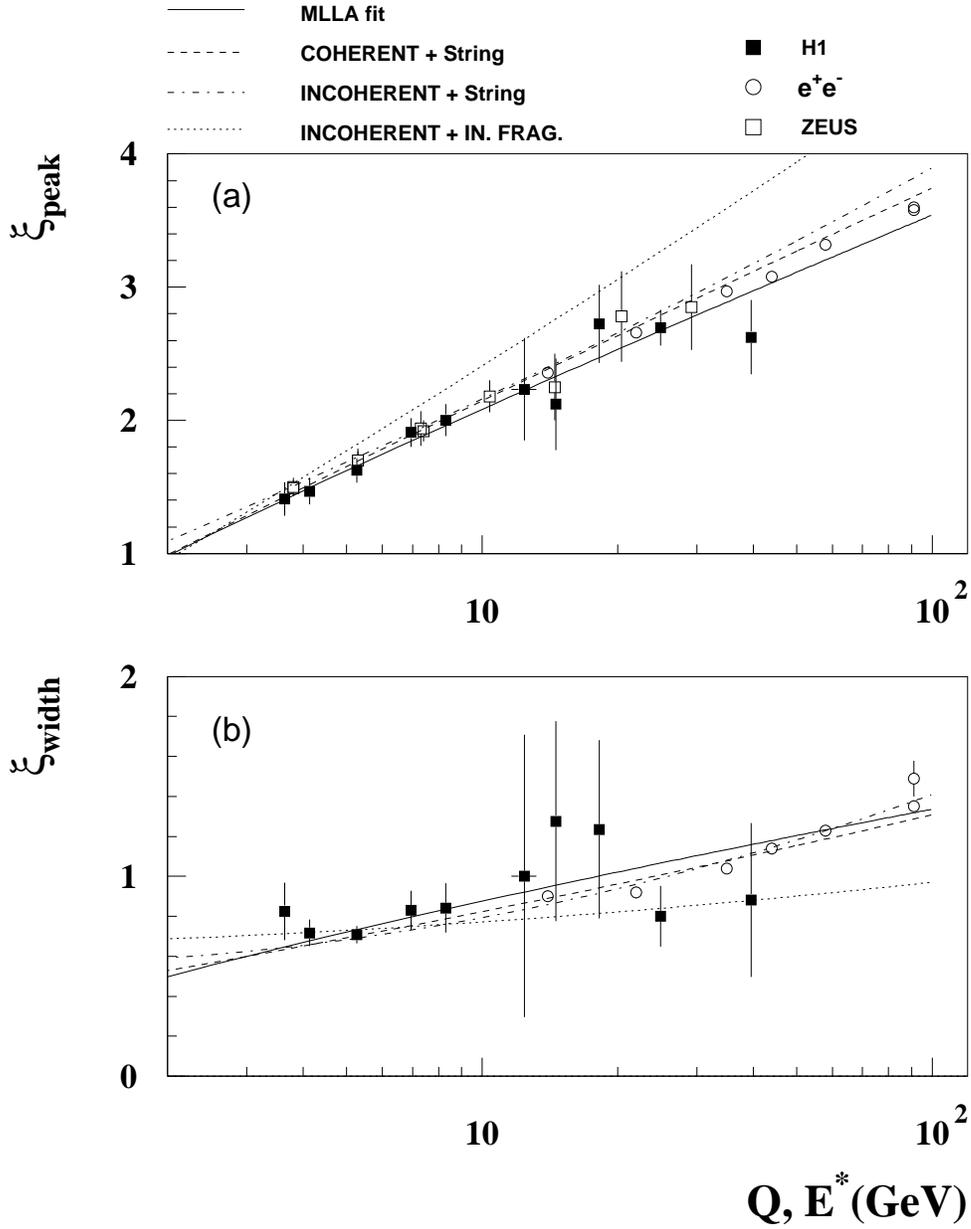,height=18.0cm}}
 \end{center}
\caption{\em Evolution of (a) the peak position and b) the width ($\sigma$) of
a fitted Gaussian fragmentation function compared with similar fits to \ee data
and with the results of a ZEUS analysis, with points at the same $Q$ being at
different $x$.
The solid line is a two-parameter simultaneous MLLA-expectation fit to the H1
data. See text for a discussion of the various model predictions.
}
\label{Pkpos}
\end{figure}

 The results of these fits for both $\xi_{peak}$ and $\xi_{width}$ variation
as a function of $Q$ are given in Table~\ref{TMult} and are shown in
Fig.~\ref{Pkpos}(a,b), with all sources of
error added in quadrature. The figure shows the results using our Breit frame
energy flow selection but, as the results in the table indicate, there is only
the suggestion of the peak values being slightly lower with the selection
rather than without. We may conclude that, within our errors,
the parameters of this analysis are insensitive to the spectrum of
hadronic fragments in QCD radiation lost to the target region and we therefore
should still be able to compare with data from \ee interactions.
The published \ee fragmentation function results [\ref{OPAL},\ref{Tasso}] have
been fitted using the same procedure and are shown at the appropriate \ee $E^*$
values. Clear agreement is again evident using $Q$ as the appropriate
equivalent variable. The peak values are also in
agreement with those of the ZEUS analysis [\ref{Zeus}] where no selection
within the current hemisphere was made.

If $Q$ is taken to be equivalent to $\sqrt{s_{ee}}$ and is normalised by an
effective energy scale to give a variable $Y=\ln (Q/2\Leff)$ then,
assuming gluon coherence, the predicted MLLA behaviour of the peak position
and width is [\ref{Dok}]
$$\xi_{peak}= 0.5Y + c_2\sqrt{Y} + O(1)$$
$$\xi_{width}= \sqrt{Y^{3\over 2}/2c_1}$$
where $c_1$, $c_2$ are constants dependent only on the number of
colours and flavours in QCD, $\Leff$ sets the scale of the mass of
the final state fragmented hadrons, and $O(1)$ contains higher order
corrections. Following [\ref{OPAL}] in assuming three flavours, we obtain
$\Leff=0.24\pm 0.04$~GeV and $O(1)=-0.38\pm 0.11$\footnote{The evolution of the
peak with $Y$ is so nearly linear that there is a large correlation between
the two parameters. This is resolved in our analysis by the width evolution
which has no dependence on the $O(1)$ term.}
in a combined fit (solid line) to the present peak and width data. These
results
are to be compared with the analysis [\ref{OPAL}] of \ee $\xi_{peak}$ evolution
[\ref{OPAL},\ref{Tasso},\ref{L3},\ref{TOPAZ}] which gives
$\Leff=0.21\pm 0.02$~GeV and $O(1)=-0.32\pm 0.06$. The dashed line of
Fig.~\ref{Pkpos} shows the prediction at the generator level of the LEPTO 6.1
Monte Carlo with an assumption of gluon coherence in parton showers and
using a Lund string model [\ref{String}] for final hadronisation.

A straight line fit to $\xi_{peak}$ as a function of
$\ln (Q)$ gives a gradient of $0.75\pm 0.05$. Reference [\ref{OPAL}] claims a
significant need for gluon coherence, noting that a gradient of unity
would be the na\"\i ve expectation if the multiplicity in a parton shower
increased solely due to the constraints of longitudinal phase space. We find
that abandoning the angular-ordering model of gluon coherence in the parton
shower, but maintaining string hadronisation (dashed-dotted line in
Fig.~\ref{Pkpos}) gives almost indistinguishable evolution predictions.
The results from HERWIG [\ref{Herwig}] (not shown) with cluster hadronisation
are similarly indistinguishable. A crude implementation [\ref{Jetset}] of an
independent fragmentation model with or without (dotted line) gluon coherence
gives effective gradients only slightly below unity ($0.94\pm 0.01$ in this
case, with our selections). However, we are aware of other implementations
[\ref{Odo},\ref{Imiph}] which impose energy conservation on hadronic
distributions of transverse momentum that reproduce the same kind of
`string' effects.
We should conclude that the common \ep /\ee result does not constitute proof of
the need for gluon coherence, and that the success of the MLLA-inspired model,
even down to our lowest $Q$ values, may be supporting claims [\ref{Boud}]
that the more fundamental cause of lower gradients is the increase of
transverse momentum with energy.


        There is an expected effect on the peak position due to the limited
opening angle of the jet implied by the selection of the current hemisphere
in the Breit frame [\ref{Dok}]. This systematically rejects lower $x_p$, and
hence higher $\xi$, tracks but there is a simultaneous inclusion of soft track
contamination from the target. As we have shown, using one hemisphere of the
Breit frame roughly equalises these effects but the investigations
[\ref{Kant}] of systematic errors which used cuts at the positions
$\cos\theta_B =\pm 0.40, \pm 0.25$ as described
earlier in this section show some evidence for a
small systematic shift in the expected direction. The net effect is small
and presently inconclusive.


\noindent  {\bf 6. Dependence on $x$} \vskip .1 true cm

\begin{figure}
 \begin{center}
  \mbox{\epsfig{file=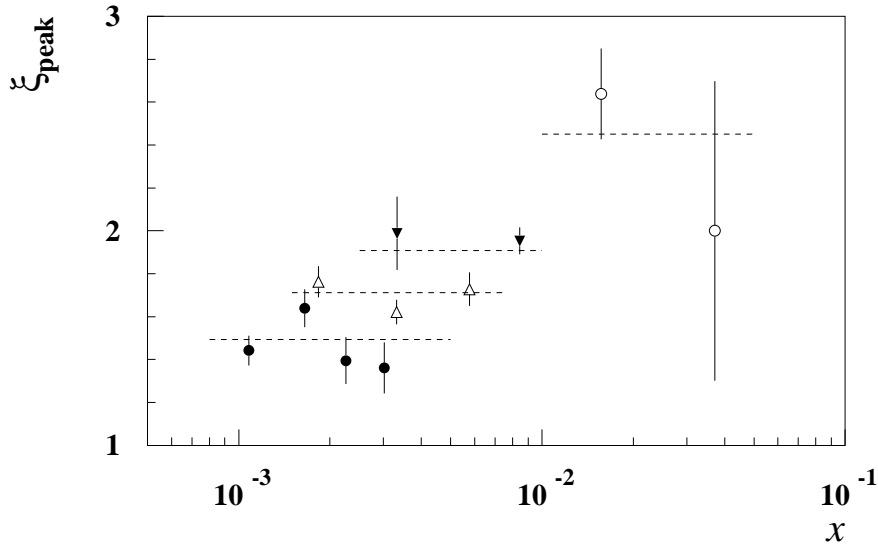,height=10.0cm}}
 \end{center}
\caption{\em {Position of the fragmentation function peak as a function of
Bjorken x, in the intervals
(solid circles) $12<Q^2<25$~{\hbox{\rm GeV}}$^2$, (open triangles)
$25<Q^2<45$~{\hbox{\rm GeV}}$^2$, (solid triangles)
$45<Q^2<80$~{\hbox{\rm GeV}}$^2$, and
(open circles) $200<Q^2<500$~{\hbox{\rm GeV}}$^2$.  The dashed lines refer to
the $\xi_{peak}$ expected from the fit at the relevant mean $Q^2$ value.
}}
\label{xdep}
\end{figure}

\noindent High $Q^2$ events are only accessible at high Bjorken $x$ and this
correlation is amplified by the $W^2$ and $y$ cuts made in this analysis.
Thus it is possible that the results presented so far are a reflected effect
of a jet fragmentation evolution in terms of the parton momentum fraction
$x$ rather than $Q^2$. The expectations of MLLA are that, although
distributions in the target region should show some $x$
dependence [\ref{Ingel}], the peak of $\xi$ in the current region should
depend only on $Q^2$. The possibility of such $x$ dependence has been
investigated, suffering the consequent loss of statistics, by defining a
number of rectangular
($x$,$Q^2$) cells well within acceptance limits and somewhat broader in
$Q^2$ than previously used. The corrected values of $\xi_{peak}$ are
displayed in Fig.~\ref{xdep} as a function of
$x$ at constant $Q^2$, with statistical errors only. The dashed lines
refer to the expected values as interpolated
from the fit to the function given in the previous section. It is clear that
these data indicate no observable $x$ dependence.

\newpage

\noindent  {\bf 7. Conclusions} \vskip .1 true cm

\noindent  This paper has shown that, when examined in the Breit frame as a
function of $Q$, the basic features of the fragmentation function of charged
particles in \ep interactions are close to those of
quarks pair-created in \ee interactions when examined in their CMS
as a function of $\sqrt{s_{ee}}$. In short, to a good approximation quarks
behave in the same way in both methods of production and $Q$ is a
suitable variable with which to describe the evolution of their fragmentation.
On the contrary there is no evidence for any dependence on $x$.
The data of this paper are consistent with the evidence from \ee data
[\ref{OPAL}] for the need to incorporate gluon coherence in the description
of this fragmentation, but Monte Carlo models {\it not} making this assumption
are also able to reproduce both sets of data quite well.

\vskip 1 cm

\noindent {\bf 8. Acknowledgements}\vskip .1 true cm


\noindent

We are grateful to the HERA machine group whose outstanding efforts made
this experiment possible. We thank the engineers and technicians for their work
in constructing and now maintaining the H1 detector, our funding agencies for
financial support, the DESY technical staff for continual assistance, and the
DESY directorate for the hospitality they extend to the non-DESY members of the
collaboration.
\vskip 1 cm

\noindent{\large \bf References}
\begin{enumerate}
\item  \label{H1spec}
       H1 Collaboration, I.\ Abt et al., \zp{63}{377}{94}; The appendix to this
publication also describes the Monte Carlo generators used in this analysis.
\item  \label{OPAL}
       OPAL Collaboration, M.Z.\ Akrawy et al., \PL{247}{617}{90}.
\item  \label{Tasso}
        TASSO Collaboration, W.\ Braunschweig et al., \zp{47}{187}{90}.
\item  \label{Hump}
        Yu.L.~Dokshitzer, V.S.~Fadin and V.A.~Khoze,
        \PL{115}{242}{82},\zp{15}{325}{82}.
        A.H.~Mueller, \NP{213}{85}{83}, \NP{241}{141}{84}.
\item  \label{Dok}
        Yu.L.~Dokshitzer, V.A.~Khoze, A.H.~Mueller and S.I.~Troyan, ``Basics of
        Perturbative QCD'' Editions Fronti\`eres (1991); Ya.I.~Azimov,
        Yu.L.~Dokshitzer, V.A.~Khoze, and S.I.~Troyan, \zp{31}{213}{86},
        \zp{27}{65}{85}; L.V.~Gribov, Yu.L.~Dokshitzer, V.A.~Khoze, and
        S.I.~Troyan, \PL{202}{276}{88},
\item  \label{Coh}
        A.H.Mueller, \PL{104}{161}{81}; A.~Bassetto, M.~Ciafaloni,
G.~Marchesini
        and A.H.~Mueller, \NP{207}{189}{82}; B.I.~Ermolaev and V.S.~Fadin,
        \JL{33}{269}{81}; Yu.L.~Dokshitzer et al., \RMP{60}{373}{88}.
\item  \label{BF}
       N.~Sakai, \PL{85}{67}{79}; J.~Sheiman, \NP{171}{445}{80};
        P.~Allen et al., \NP{176}{333}{80}.
\item  \label{GTetal}
       G.\ Thompson et al., \JP{19}{1575}{93}.
\item  \label{Char}
       K.\ Charchula, \JP{19}{1587}{93}.
\item  \label{Ingel}
       G.\ Ingelman and J.\ Rathsman, \JP{19}{1594}{93}.
\item  \label{H1det}
        H1 Collaboration, I.\ Abt et al., DESY preprint 93-103 (1993),
submitted
to Nucl. Instr. Meth.
\item  \label{H1CDT}
        J.\ B\"urger et al., \NIM{279}{217}{89}.
\item  \label{FTD}
        H1 FTD group, S.Burke et al., submitted to Nucl. Instr. Meth.
\item  \label{GEANT}
        R.~Brun et al., GEANT3 User's Guide, CERN preprint
        CERN-DD/EE~84-1.(1987)
\item  \label{H1CAL}
        H1 calorimeter group, B.~Andrieu et al., \NIM{336}{460}{93}.
\item  \label{diff}
        H1 Collaboration, I.~Abt etal., \NP{429}{477}{94}.   
\item  \label{CDM}
        L.~L\"onnblad, \CPC{43}{367}{92}.
\item  \label{Lepto}
        G.~Ingelman, Proc. HERA workshop, W.~Buchm\"uller, G.~Ingelmann (eds.),
        Hamburg (1991) vol.3, 1366;
         H.~Bentsson, G.~Ingelmann, and T.J.~Sj\"ostrand, \NP{301}{554}{88}.
\item  \label{Djn}
        G.A.\ Schuler and H.\ Spiesberger. Physics at HERA Workshop
Vol.3, (1991) 1419.
\item  \label{eemult}
        OPAL collaboration, P.D.~Acton et al., \zp{53}{539}{92}.
\item  \label{Zeus}
        ZEUS Collaboration, DESY preprint DESY 95-007, 1995.
\item  \label{Delphi}
        DELPHI Collaboration, P.~Abreu et al., \PL{347}{447}{95}.
\item  \label{EMC1}
        EMC, M.\ Arneodo et al., \NP{258}{249}{85}.
\item  \label{EMC665}
        EMC, M.\ Arneodo et al., \zp{35}{417}{87};
E665 Collaboration, \PL{272}{163}{91}.
\item  \label{L3}
        L3 Collaboration, B.\ Adeva et al., \PL{259}{199}{91}.
\item  \label{TOPAZ}
        TOPAZ Collaboration, R.\ Itoh et al., \PL{345}{335}{95}.
\item  \label{String}
        B.~Andersson, G.~Gustafson, G.~Ingelman and T.~Sj\"ostrand,
\PR {97}{31}{83}; T.~Sj\"ostrand, \NP {248}{469}{84}.
\item  \label{Herwig}
        G.~Marchesini et al., \CPC{67}{465}{92}.
\item  \label{Jetset}
        T.J.~Sj\"ostrand, CERN preprint CERN-TH.7112/93 and \CPC{43}{367}{87}.
\item  \label{Odo}
        P.~Mazzanti and R.~Odorico, \NP{370}{23}{92}.
\item  \label{Imiph}
        P.~Biddulph and G.~Thompson, \CPC{54}{13}{89}.
\item  \label{Boud}
        E.R.~Boudinov, P.V.~Chliapnikov and V.A.~Uvarov, \PL{309}{210}{93}.
\item  \label{Kant}
        D.~Kant, University of London thesis, to be published.
\end{enumerate}
\end{document}